\documentclass[aps,prl,twocolumn,showpacs,superscriptaddress,floatfix]{revtex4-1}

\usepackage{graphicx}
\usepackage{amsmath,amssymb,latexsym}
\usepackage{color}
\bibstyle{apsrev.bib}

\newcommand{\mean}[1]{{\left \langle {#1} \right \rangle}}

\begin{document}
\def\be{\begin{equation}}
\def\ee{\end{equation}}

\def\bfi{\begin{figure}}
\def\efi{\end{figure}}
\def\bea{\begin{eqnarray}}
\def\eea{\end{eqnarray}}

\title{Nonequilibrium fluctuation-dissipation theorem and heat production}
\author{E. Lippiello}
\email{eugenio.lippiello@unina2.it}
\affiliation{Department of Mathematics and Physics, Second University of Naples, Via Vivaldi 43, 81100 Caserta, Italy}
\author{M. Baiesi}
\affiliation{Department of Physics and Astronomy, University of Padua, Via Marzolo 8, I-35131 Padova, Italy}
\affiliation{INFN, Sezione di Padova, Via Marzolo 8, I-35131 Padova, Italy}
\author{A. Sarracino}
\affiliation{ISC-CNR and Dipartimento di Fisica,
Universit\`a ``Sapienza'', p.le A. Moro 2, 00185 Rome, Italy}
\affiliation{Laboratoire de Physique Th\'eorique de la Matiere Condens\'ee, CNRS UMR 7600,
case courrier 121, Universit\'e Paris 6, 4 Place Jussieu, 75255 Paris Cedex}

\date{\today}

\begin{abstract}
We use a relationship between response and correlation function in
nonequilibrium systems to establish a connection between the heat
production and the deviations from the equilibrium
fluctuation-dissipation theorem.  This scheme extends the Harada-Sasa
formulation [Phys. Rev. Lett. 95, 130602 (2005)], obtained for
Langevin equations in steady states, as it also holds for transient regimes and for 
discrete jump processes involving small entropic changes.
Moreover, a general formulation includes two times and the new
concepts of two-time work, kinetic energy, and of a two-time heat
exchange that can be related to a nonequilibrium ``effective
temperature''. Numerical simulations of a chain of anharmonic
oscillators and of a model for a molecular motor driven 
by ATP hydrolysis illustrate these points.
\end{abstract}


\maketitle

For systems in equilibrium, the response $R$ to a small external perturbation 
is related to the correlation function $C$ of spontaneous fluctuations. 
This property is described by the fluctuation-dissipation theorem (FDT).
In the last two decades, many efforts have been devoted to the study and the
understanding of deviations from FDT in nonequilibrium
systems~\cite{CR03,BPRV08,SS10,BM13}.  These studies can be
substantially split in two research lines: i) establishing a
connection between deviations from FDT and thermodynamic properties;
ii) searching for general relations including $R$ and $C$ that hold
also out of equilibrium.

Along the first line there is the definition of an
effective temperature in nonequilibrium systems, 
based on the formula of the equilibrium FDT and 
coinciding with the temperature $T$ of the thermal bath if the FDT holds~\cite{CKP97}.
The concept of effective temperature
has been introduced in a variety of systems including aging
or driven systems and quantum quenches~\cite{leticia}.  
More recently Harada and Sasa (HS) have derived a relation between the average rate of energy
dissipation $J$ and deviations from the FDT~\cite{HS05,HS06,H09}, for systems described by a
Langevin equation with white noise (see also the extension to colored noise~\cite{DN06}).
Indicating with $C$ the velocity correlation function and with $R$ the
change of velocity caused by a constant external force, the HS relation reads
\begin{equation}
\mean{J} = \gamma \int_{-\infty}^{\infty} [\tilde C(\omega) - 2 T \tilde R_S(\omega)]
\frac{d\omega}{2\pi} ,
\label{eq1}
\end{equation}
where $\langle \dots\rangle$ denotes an average over trajectories.
The friction coefficient  $\gamma$ is related to the typical
inverse time of energy dissipation. Quantities in Fourier space are
denoted with a tilde, and $\tilde R_S(\omega)$ is the transform of the symmetric
part $R_S(t)$ of the response function (i.e.~the real part of the transform).
The advantage of this approach is that
correlation functions and response functions of fluctuating
observables are often more easily accessible in experiments and one
can thus utilize deviations from FDT to indirectly infer the rate of
energy dissipation (see also a recent alternative~\cite{LMBBS12}).
For example, it was utilized to study optically driven colloidal 
systems~\cite{TJNMS07}  and the energetics of a model for molecular motor~\cite{TOWTKM10}.
Although it was originally obtained for nonequilibrium steady states, 
the formula~(\ref{eq1}) was also used in the microrheology of a particle trapped 
in a relaxing lattice~\cite{GSPC12}. 

Concerning the research line (ii), many expressions of $R$ in terms of
quantities of the unperturbed dynamics have been proposed in the last
years for a wide range of nonequilibrium systems (see for
example~\cite{CKP94,LCZ05,ss06,BMW09,GPCCG09} and references
in~\cite{CR03,BPRV08,SS10,BM13}).  In this letter we will focus on the
following generalized FDT (GFDT)~\cite{LCZ05,BMW09}.  Indicating with
$R=\partial \mean {Z(t_2)}/\partial h(t_1)\vert_{h=0}$ the response of
an observable $Z$ to an external perturbation $h$
coupled with the observable $Y$, and fixing $t_2 \ge t_1$, the GFDT
reads 
\be 2 T R(t_1,t_2)=\frac{\partial \mean{Z(t_2)Y(t_1)}}{\partial
  t_1}-\mean{Z(t_2)B(t_1)}
\label{eq2}
\ee
where the observable $B$ satisfies the relation
\be
\mean{B(t_2)Z(t_1)}= \frac{ \partial \mean{Y(t_2)Z(t_1)}}{\partial t_2}
\label{eq3}
\ee 
for any $Z$. This GFDT holds for systems described by Markovian dynamics,
under quite general assumptions~\cite{LCZ05,BMW09}.

In this Letter we present a connection between the GFDT~(\ref{eq2}) and
the HS relation~(\ref{eq1}).
More precisely we prove that a relation between 
heat flux, response function, and velocity correlation
holds a) not only for steady states but in
general transients; b) for general discrete variables evolving according to 
jump processes, such as particle collisions or the positions in a discrete space; 
c) not only on average but also for suitably defined fluctuating parts of $C$, $R$ and $J$.  Our
framework includes novel quantities: a ``two-time'' generalized heat,
a ``two-time'' kinetic energy and a ``retarded-anticipated'' work. The
GFDT allows one to find a connection between these ``two-time''
observables and eventually to support the definition of an effective
temperature from FDT deviations in systems where timescales 
corresponding to different degrees of freedom are well-separated.

We consider a system of $N$ particles and indicate with 
${\bf x}_i(t), {\bf v}_i(t)$ the position and velocity of the $i$-th particle 
at time $t$ and with $m_i$ its mass. 
Each particle is affected by a force ${\bf f}_i(t)$ that is the sum of internal and external
interactions and that can be non-conservative. 
We first develop an approach where time is discretized with fixed time step $\tau$ and 
the local velocity $v_i(t)$ can be updated to a new value $v'_i(t)$ 
with a stochastic evolution characterized by transition rates
$\Omega (v_i(t) \to v'_i(t))$, while positions follow deterministically
$x_i(t+\tau)=x_i(t)+ v'_i(t) \tau$.
Such approach can reproduce Langevin inertial dynamics.
Later we will shift the attention to discrete-state systems where positions change stochastically
and velocities are subordinate variables.
Within this formalism time derivatives are discrete:
$\tau \partial v_i(t)/\partial t=\delta v_i(t)$ and $\tau \partial x_i(t)/\partial t=\delta x_i(t)$.
To simplify the notation hereafter we restrict to a one-dimensional motion and we drop the
particle index, the generalization to more variables being trivial for quantities with
uncorrelated fluctuations.

A ``two-time'' kinetic energy and an elementary retarded-anticipated work~\cite{ZBCK05} 
during $\tau$  are defined as 
\bea 
\hat K(t_1,t_2) &\equiv& \frac{m}{2} v(t_2) v(t_1) \label{k}\\
\delta \hat W(t_1,t_2) &\equiv& \frac{1}{2} \tau \left [f(t_2)v(t_1)+v(t_2)f(t_1) \right] \,. \label{W}
\eea 
The notation $\hat X(t)$ is useful to explicitly indicate fluctuating quantities,
namely $X(t)=\langle \hat X(t)\rangle$ (however, positions and velocities are
always meant to be fluctuating). 
In the  $t_2\to t_1$ limit, $\hat K$ and $\hat W$  coincide with the
standard definition of kinetic energy and work. Moreover, for an isolated system not
coupled with a thermal reservoir ---  where $f(t)=m\ \delta v(t)/\tau$  --- it is easy to show
that they satisfy a Work-Energy theorem for arbitrary $t_2$ and $t_1$,
\be
\delta \hat W(t_1,t_2)=\frac{m}{2}\left [\delta v (t_2)v(t_1)+ v (t_2)
\delta v(t_1) \right]= d \hat K(t_1,t_2) .
\ee
It is therefore natural to define the ``two-time heat exchange'' (from the reservoir to the system) as
\be
\delta \hat Q(t_1,t_2)\equiv d \hat K(t_1,t_2) - \delta \hat W(t_1,t_2).
\label{Q}
\ee 
The two-time kinetic energy is trivially proportional to the
velocity-velocity fluctuations, namely $\hat C(t_1,t_2)\equiv v(t_2)
v(t_1)$.  Notice that our definition of heat is related to Eq.~(59) of
Ref.~\cite{HS06} (written for Langevin equations at stationarity), 
and in the limit of equal times it has the structure of the heat
defined in stochastic energetics~\cite{sekimoto}.
One can express $\delta \hat W$ in terms of the velocity correlation function $\hat C(t_1,t_2)$ and of 
a ``fluctuating'' response function $\hat R(t_1,t_2)$ of the velocity $v(t_2)$ to an
external force switched on during the microscopic time interval 
$[t_1,t_1+\tau]$ (see below). 

Observing that the external force $h(t_1)$ perturbs the system
energy by a factor $h (t_1) v(t_1)\tau$,
the response function $R(t_1,t_2)$ can be immediately obtained putting $ Z(t_2)=v(t_2)$
and $Y(t_1)=v(t_1) \tau$ in Eqs~(\ref{eq2},\ref{eq3}), which for discrete time dynamics read
\bea 2 T
R(t_1,t_2) &=& \mean{v(t_2)\delta v(t_1)} - \mean {v(t_2) \hat B(t_1)}  \label{eq4} \\
\mean{\hat B(t_2) v(t_1)} &=&  \mean{\delta v(t_2)  v(t_1)}.            \label{eq5}
\eea
The next step is to express $\hat B$ in terms of system properties. 
For simplicity, in this derivation we allow variations of $v \to v'=v\pm\epsilon$.
Since velocities evolve with jumps, we have \cite{LCZ05,BMW09}
\be 
\hat B(t)=\epsilon \tau \left[\Omega(v \to v+\epsilon)-\Omega(v\to v-\epsilon)
\right ]
\label{bt}
\ee
and, assuming local equilibrium, transition rates $\Omega(v \to v')$ embody the property of
local (or generalized) detailed balance~\cite{LS99,CM99},
\be
\Omega (v \to v+\epsilon) =\alpha(\epsilon) e^{[dS(v+\epsilon)-dS(v)]/2}, 
\label{n8}
\ee   
where $dS(v+\epsilon) - dS(v)$ is the additional entropy change in the environment caused by the velocity jump and 
$\alpha(\epsilon)$ is a symmetric function of $\epsilon$ fixing the jump rates. 
For systems in contact with a single heat bath with $\beta=1/T$ (setting the Boltzmann constant $k_B=1$),
the entropy change should take the form
$dS(t) = - \beta\, \delta \hat Q(t,t)$ with the heat transferred 
defined in Eq.~(\ref{Q}) leading to
\be
dS(v+\epsilon) - dS(v) = - \beta \epsilon [ m v   - f \tau ]. 
\ee
With this equation, and 
assuming that $\zeta\equiv\beta m \epsilon^2 \ll 1$ one expands the exponential in Eq.~(\ref{n8}) to find
$\hat B(t)= -\alpha(\epsilon) \zeta \tau [v(t) - \frac \tau m  f(t)]$.
Therefore fixing the jump rates such that $\alpha(\epsilon) \zeta \tau=1$ 
one obtains
\be
\hat B(t)  =-v(t) +  \frac{\tau}{m} f(t). 
\label{n7}
\ee 
The master-equation with transition rates~(\ref{n8}), in the limit of
$\epsilon \to 0$, may be set up to converge to a Fokker-Planck
equation with drift term given by Eq.~(\ref{n7}) and diffusion
coefficient $D = \frac{\gamma T}{m^2}$~\cite{vK61,SVCP10}.
%In the limit $\zeta \ll 1$ the first three moments of the transition rates
%$M_n=1/n!\sum_{v'}(v'-v)^n\Omega(v \to v')$ give respectively
%$M_0=2\alpha$, $M_1=\hat B(t)/\tau$, and $M_2=\alpha \epsilon^2$.
%Expressing time in units of $\tau \zeta$, in the limit $\zeta \to 0$,
%time becomes continuous and a suitable convergence in the $\epsilon
%\to 0$ limit~\cite{G90} is obtained under the scaling behaviors
%$\alpha= D\epsilon^{-2}$ and $\tau\zeta=D^{-1}\epsilon^2$.  
In this limit the friction coefficient becomes $\gamma \equiv m / \tau$.

The insertion of Eq.~(\ref{n7}) in~(\ref{eq4},\ref{eq5}) 
leads to an interesting physical structure of the FDT,
$2 T R(t_1,t_2) = \mean{v(t_2) d {\Phi}(t_1)} - \mean{v(t_2) d {\Psi}(t_1)}$, in which
a time-antisymmetric dissipative term $d {\Phi}(t_1) = v(t_1)$, coming from Eq.~(\ref{n7}), 
is associated with the well known concept of entropy production~\cite{BMW09,BM13}, 
here in excess due to the perturbation.  
The time-symmetric term $d {\Psi}(t_1) = \tau f(t_1)/m - \delta v(t_1)  = - a_{\rm{bath}}(t_1)\tau$ 
in this case is related to the acceleration $a_{\rm{bath}}$ generated by the thermal bath in the time interval
$\tau$. See for instance Refs.~\cite{lec05,mer05,mae06,BM13} for recent discussions on time-symmetric observables. 

To conclude our argument, the final step is to combine
terms of the FDT in a form enjoying the invariance of the generalized heat exchange 
$\delta \hat Q(t_1,t_2)$ for swaps of $t_1$ with $t_2$.
We define a fluctuating response function 
$\hat R(t_1,t_2) \equiv [v(t_2)\delta v(t_1) - v(t_2) \hat B(t_1)] / (2T)$, 
whose statistical average satisfies (\ref{eq4}), and we consider its symmetric part
\bea
&& \hat R_S(t_1,t_2)\equiv [\hat R(t_1,t_2)+\hat R(t_2,t_1)]/2 = \label{Rs}  \\ 
&&=\frac { v(t_2)\delta v(t_1)+\delta v(t_2) v(t_1) - v(t_2)\hat B(t_1)-\hat B(t_2) v(t_1)}{4 T} .
\nonumber
\eea
Its average is $R_S(t_1,t_2)=\frac 1 2 R(t_1,t_2)$ for $t_2>t_1$ because
$R(t_2,t_1)=0$ for causality. 
From Eq.~(\ref{Q}) and by using Eq.~(\ref{n7}) in~(\ref{Rs}) one obtains
\be
{\delta \hat Q(t_1,t_2)}=m \left [2 T \hat R_S(t_1,t_2)- \hat C(t_1,t_2) \right ],
\label{dQ}
\ee
valid for any single trajectory. Its average over trajectories yields a two-time generalized HS relation,
\be 
{J(t_1,t_2)}=\gamma \left [2 T  R_S(t_1,t_2)- C(t_1,t_2) \right ],
\label{J}
\ee 
including a generalized heat production rate $J(t_1,t_2)=\delta Q(t_1,t_2)/\tau$.
Eq.~(\ref{J}) expresses heat production in terms of  
deviations from the equilibrium FDT $T R(t_1,t_2)=C(t_1,t_2)$.
This relation can be directly derived by time-reversing Eq.~(\ref{eq5}) and
using the result in (\ref{eq4}) to keep only the terms with velocities. 
Therefore, $J(t_1,t_2)$ is identically null in equilibrium. 
Notice that the HS relation~(\ref{eq1}), or its generalization Eq.~(58) 
of Ref.~\cite{HS06}, are readily recovered by assuming a
steady state (such that two-time quantities are function only of the
time differences $t_2-t_1$) and taking the Fourier transformation of
Eq.~(\ref{J})~\footnote{In the original HS formula the correlation $C$ was computed for
velocities after subtracting their steady state mean value, a step not necessary
in our derivation}.

Eq.~(\ref{J}) can be obtained also for the overdamped regimes~\cite{HS06}. 
Now positions change stochastically according to a master equation, with jumps $x(t+\tau)=x(t)\pm \lambda$,
and the external force $h(t_1)$ perturbs the system energy by a factor $h (t_1) \delta x(t_1)$.
Setting $ Z(t_2)=\delta x(t_2)/\tau$ and $Y(t_1)=x(t_1)$  in Eqs~(\ref{eq2},\ref{eq3}),
one has
\be 2 T
R(t_1,t_2)= \mean{\frac{\delta x(t_2)}{\tau}\frac{\delta x(t_1)}{\tau}} -\mean {\frac{\delta x(t_2)}{\tau} \hat B(t_1)} 
\label{eq4b}
\ee
with $\mean {\hat B(t) }=\mean{\delta x(t)}/\tau$.
Since the kinetic energy is null,
one has $\delta \hat Q(t_1,t_2)= \delta \hat W(t_1,t_2)$ from Eq.~(\ref{Q}).
Eq.~(\ref{n8}) now becomes
\be
\Omega (x \to x'=x\pm \lambda) =\alpha e^{[S(x')-S(x)]/2}, 
\label{n8bis}
\ee 
with $S(x')-S(x) = - \beta\, \delta \hat Q = \mp \beta f(x) \lambda$.
Assuming $\beta \lambda \hat Q \ll 1$, one obtains
$\hat B(t)=\alpha\beta \lambda^2 f[x(t)]= f[x(t)]/\gamma$ with $1/\gamma=\alpha\beta\lambda^2$, and
using Eq.~(\ref{eq4}) one recovers the 
HS relation~(\ref{J})~\footnote{To symmetrize $R$ at $t_2=t_1$, 
one considers a contribution for $B$ before the transition and one following it.
The correct form thus involves 
$\delta \hat Q(t,t) = \frac{1}{2} [B(t)+B(t+\tau)] [x(t+\tau)-x(t)]$, which yields
the correct Stratonovich convention for $\delta Q$ if the limit to diffusion processes is permormed.
}.

In this context one can broaden the view by considering variables 
$x$ that are not positions, e.g.~they can be chemical levels.
The limit of applicability of the previous formulas is given by the constraint $\beta \lambda \hat Q\ll 1$.
If it is not satisfied, the replacement of terms $\sim \exp(S)$ with $1+S$ in the
equations for $B$ is not justified. Hence, the formalism we have developed works for systems evolving with
discrete jumps, provided that these jumps do not change the environment's entropy substantially. 

Eq.~(\ref{dQ}) may support the definition of an effective temperature as
 $T_{\textrm{eff}}(t_1,t_2) =\frac { C(t_1,t_2)}{ 2 R_S(t_1,t_2)}$~\cite{CKP97}.  
From this definition and from Eq.~(\ref{dQ}), averaging over trajectories one may write
$\delta Q(t_1,t_2)= 2 m  R_S(t_1,t_2) [T - T_{\textrm{eff}}(t_1,t_2) ]$, 
to emphasize that the heat flux depends on the
difference between the local (in time and space) temperature 
$T_{\textrm{eff}}$ and the bath temperature. 
In the limit $t_2 \to t_1$ we have $m R(t_1,t_2) \to 1$ 
(or $2 m R_S(t_1,t_2) \to 1$), hence
\be
{\delta Q(t_1,t_1)}=\left [T - m \langle v(t_1)^2 \rangle \right ] .
\label{dQ3}
\ee

%%%%%%%%%%%%%%%%%%%%%%%%%%%%%%%%%%%%%%%%%%%%%%%%%%%%%%%%%%%%%
\begin{figure}[!t]
\begin{center}
%\includegraphics[width=.9\columnwidth,clip=true]{5sfere.eps}
%\vskip 0.5cm
\includegraphics[width=.8\columnwidth,clip=true]{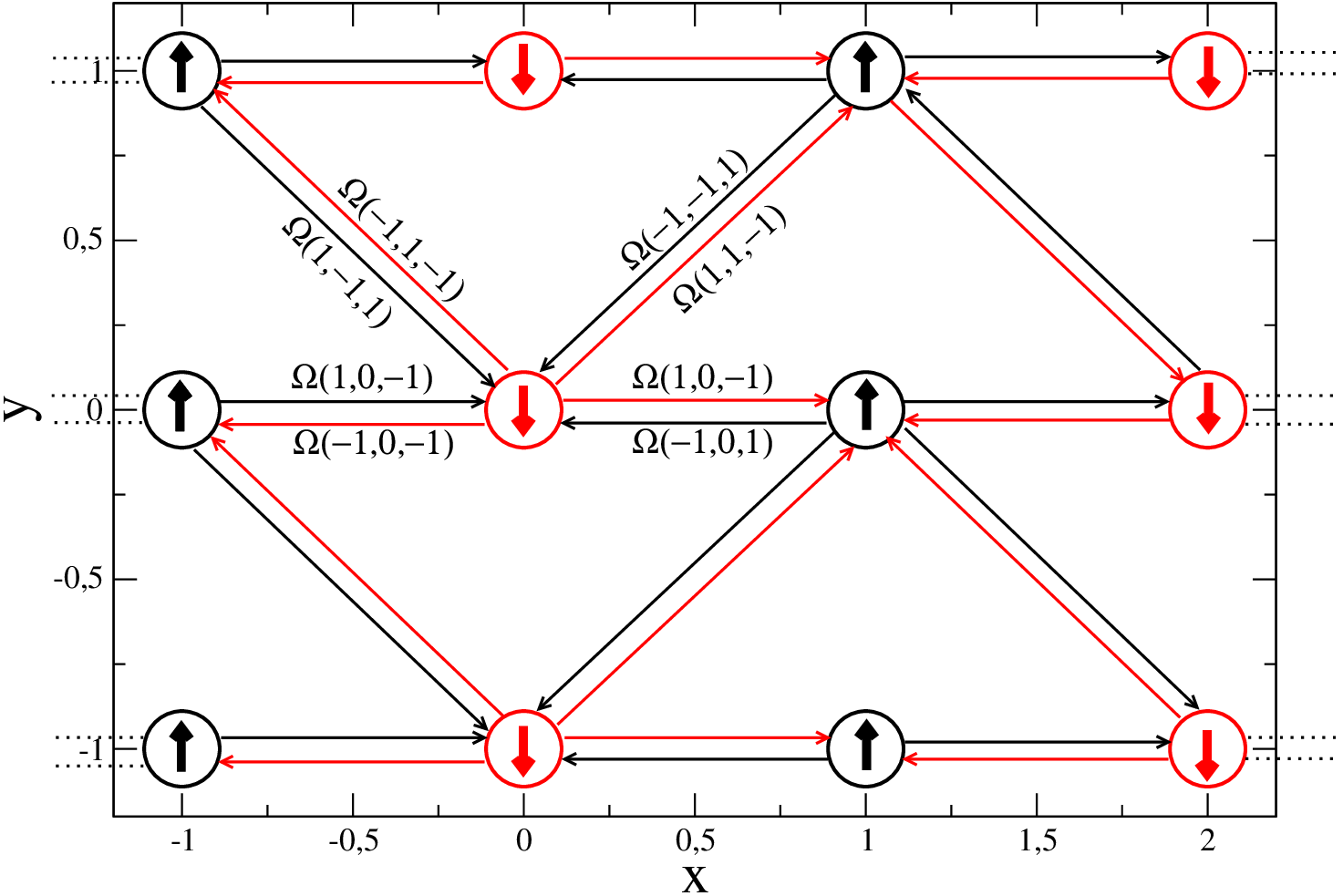}
\end{center}
\caption{(Color online)
A schematic of transition rates for the molecular motor corresponding to model (ii). Thick arrows 
indicate configurations with $\sigma$ up or down.
} 
\label{fig1}
\end{figure}
%%%%%%%%%%%%%%%%%%%%%%%%%%%%%%%%%%%%%%%%%%%%%%%%%%%%%%%%%%%%%

We illustrate the above results in two simplified systems: 
i) a chain of coupled non-linear oscillators with Hamiltonian
$H=\frac{1}{2}\sum_{i=0}^{N} \left (m  v_i^2+A_1 \left ( x_{i+1}- x_i\right )^2+
\frac{2}{3} A_2 \left ( x_{i+1}- x_i\right )^4 \right)$, 
with fixed boundary conditions $ v_0= v_{N+1}=x_0= x_{N+1}=0$
where velocities evolve according to the Metropolis algorithm; 
ii) a model for a molecular motor driven by ATP hydrolysis introduced in~\cite{LLM07}. 
In the latter model the evolution corresponds to an overdamped dynamics 
in two dimensions with transition rates for a jump
$\delta x=0,\pm 1$ and $\delta y=0,\pm 1$, along the $x$ and $y$ direction respectively, 
 given by $\Omega (\delta x,\delta y,\sigma)=\alpha \Theta(\delta x,\delta y,\sigma)
e^{K\delta x \sigma + \frac{\varepsilon}{2}\sigma + \frac{\Delta \mu}{2}\Delta y}$.
Here $\sigma=\pm 1$ indicates the states $b$ and $a$ of~\cite{LLM07}, respectively, 
and $\Theta(\delta x,\delta y,\sigma)$ is a kinetic constraint that confines trajectories along the paths 
represented in Fig.~\ref{fig1}. 
We set the parameter $\varepsilon$, corresponding to heat exchange by thermal activation, 
to zero and define $\delta x'=\sigma \delta x$. This choice allows us to decouple the 
variables fluctuations, and in this case the evolution simply corresponds to discrete jumps 
in the $x'-y$ plane,
with kinetic constraints, under the action of an external force with components $\vec f=(K,\Delta \mu/2)$ 
and $\gamma=1$.  
As a consequence, Eq.~(\ref{J}) is still recovered by simply generalizing the above arguments to a two 
dimensional evolution.

%%%%%%%%%%%%%%%%%%%%%%%%%%%%%%%%%%%%%%%%%%%%%%%%%%%%%%%%%%%%%
\begin{figure}[!t]
\begin{center}
\includegraphics[width=.8\columnwidth,clip=true]{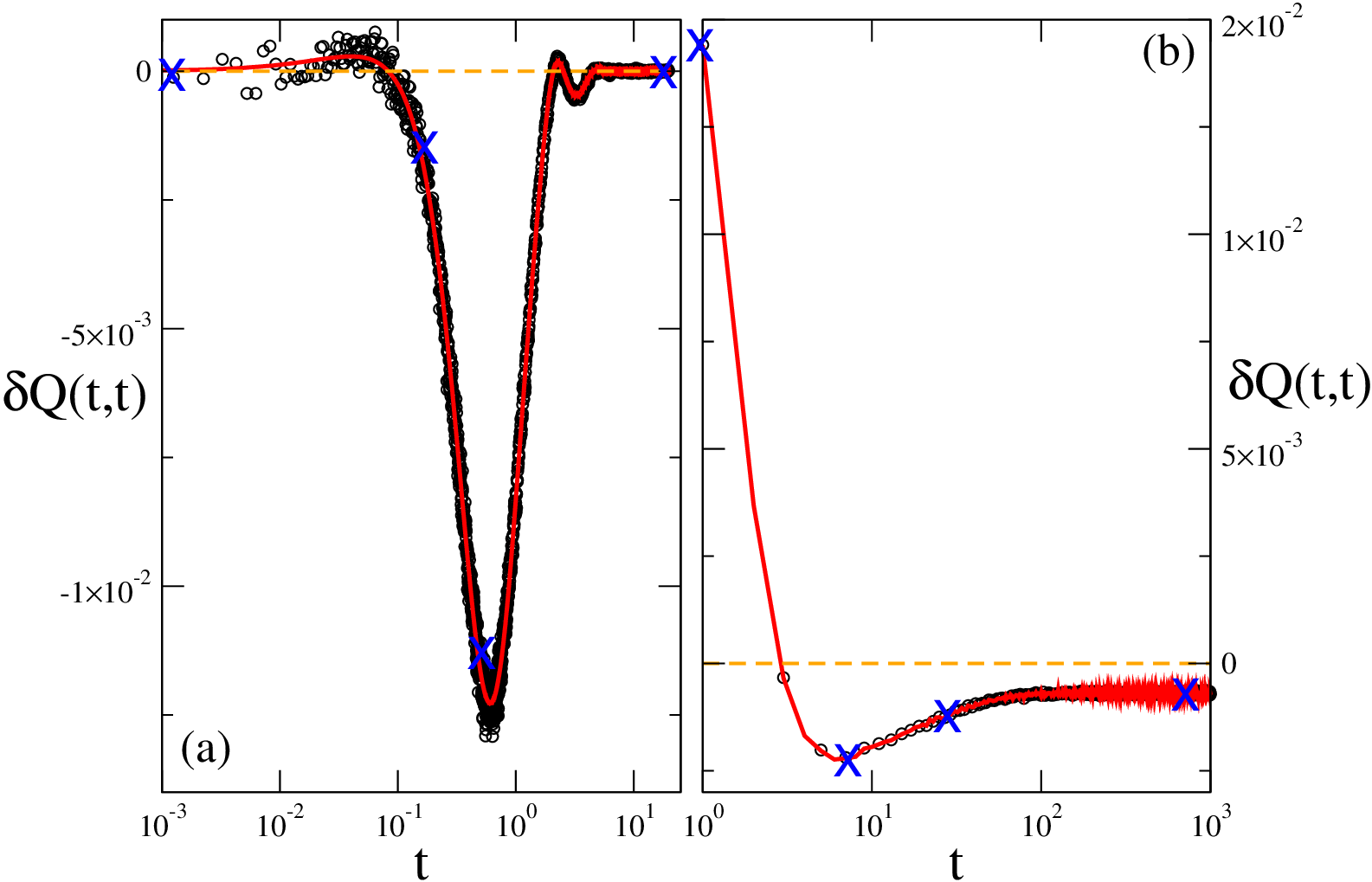}
\end{center}
\caption{ 
(Color online) The average heat exchange in (a) model (i) and (b) model (ii).  
 Black circles are direct estimates obtained from Eq.~(\ref{Q})
 ($10^7$ data).  The red continuous line is the estimate obtained by averaging Eq.~(\ref{dQ}). 
Blue crosses indicate the values of $t_1$ used in the next figure.}
\label{fig2}
\end{figure}
%%%%%%%%%%%%%%%%%%%%%%%%%%%%%%%%%%%%%%%%%%%%%%%%%%%%%%%%%%%%%

In Fig.~\ref{fig2} we plot the average value of the equal-time heat-exchange
$\delta Q(t,t)$ for both models. In model (i) the temperature $T$, mass $m$, and $\zeta$ are fixed, 
time is measured in unit of Monte Carlo steps multiplied by $\zeta$
and energy is measured in units of the temperature $T$. We show results
for a system with $N=32$, $T=100$, $m=10$, $\zeta=5\cdot 10^{-4}$,  $A_1=800$ and
$A_2=80$.
In model (ii) we set  $K=0.05$, $\Delta \mu=0.2$ and $\zeta_{\lambda}=1$ 
with initial position $x=0$, $y=0$ and $\sigma=-1$. 
Other parameters or initial conditions yield similar results.
In both cases $\delta Q(t,t)$ is a
non-monotonic function of time with both positive and negative values
and relaxing to a the equilibrium value $\delta Q(t,t)=0$ for $t>t_{st} \simeq
5$ in model (i) and to a stationary asymptotic value $\delta Q_{\infty}$ smaller than zero in model (ii) with $t_{st} \simeq 200$.  
We wish to stress that Eq.~(\ref{J}) is verified for both models in the whole temporal range.

\begin{figure}[!t]
\begin{center}
\includegraphics[width=0.98\columnwidth,clip=true]{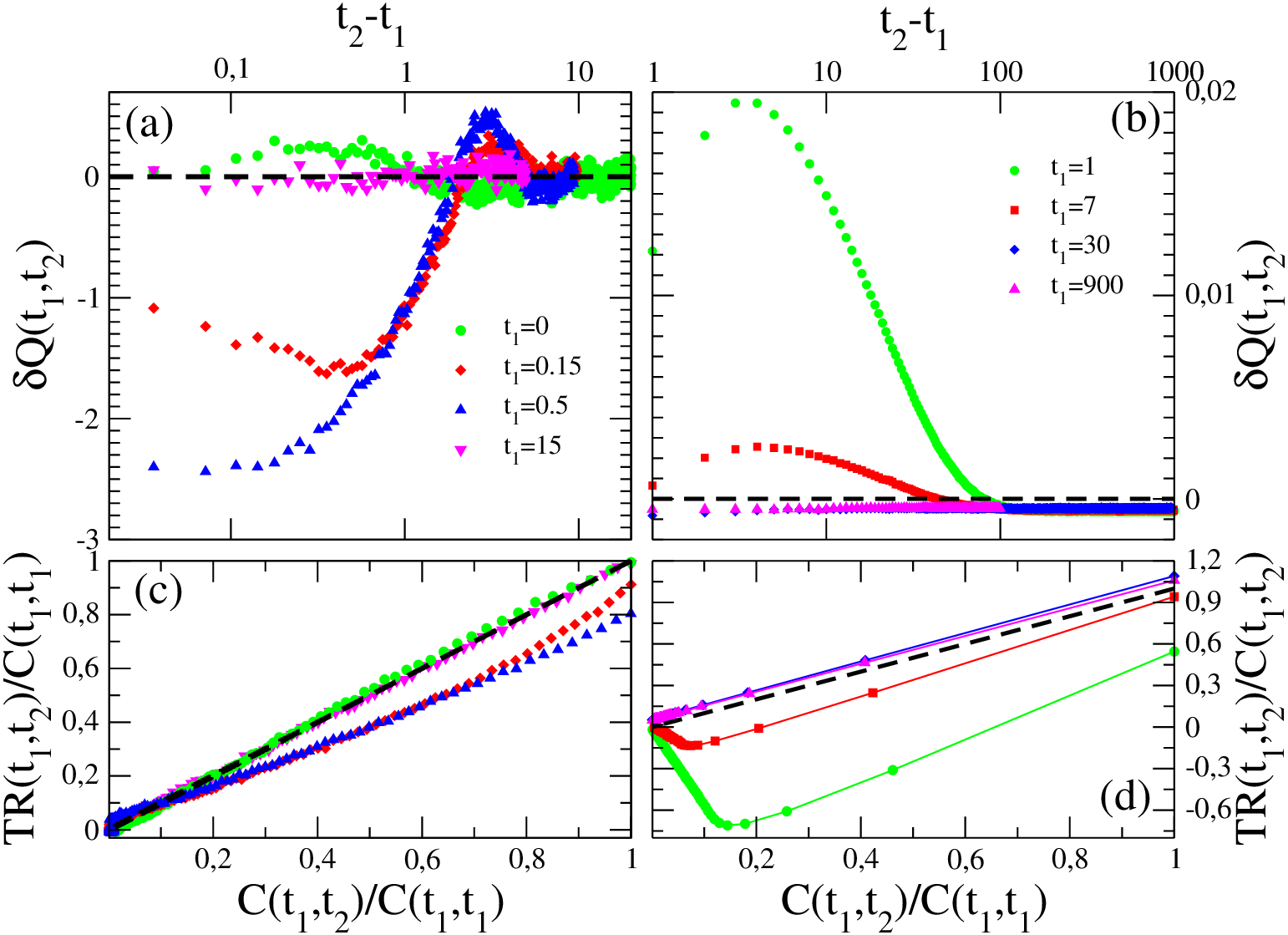}
\end{center}
\caption{(Color online)
The upper panels show the generalized heat vs.\ time, calculated with Eq.~(\ref{dQ}) 
for (a) model (i) and (b) model (ii).
The lower panels show the parametric plot of the response function vs. correlation function (both rescaled
by $1/C(t_1,t_1)$) for (c) model (i) and (d) model (ii). 
The diagonal (black  line) represents the equilibrium FDT in the form $TR=C$.}
\label{fig3}
\end{figure}

In the upper panels of Fig.~\ref{fig3} we plot  $\delta Q(t_1,t_2)$ as function of $t_2-t_1$ for different values of $t_1$ corresponding to blue crosses in Fig~\ref{fig1}. We observe that for model (i) (panel a) $\delta Q(t_1,t_2)$ is a non monotonic function with positive and negative values that converges to zero asymptotically ($t_2-t_1 \gg 1$). 
 Recalling the relationship between $\delta Q(t_1,t_2)$ and $T_{\textrm{eff}}$, this figure indicates that the effective temperature oscillates around the bath temperature $T$. The amplitude of the fluctutaions are controlled by the initial value of 
$\delta Q(t_1,t_1)$. In particular when $t_1>t_{st}$, equilibrium is reached and $\delta Q(t_1,t_2)=0$ in the whole temporal range. This is confirmed by the lower panel (panel c) where we  plot the response function $R(t_1,t_2)$ 
versus the two-time correlation function $C(t_1,t_2)$
for different values of $t_1$. 
The equilibrium condition, corresponding to a straight line with slope $1$, is recovered for $t>t_{st}$. Furthermore, larger values of $\vert \delta Q(t_1,t_1)\vert$ 
also correspond to larger deviations from the equilibrium FDT. 
Also for model (ii) $\delta Q(t_1,t_2)$ presents both positive and negative values (Fig.~\ref{fig3}b) with amplitudes controlled by the initial value $\delta Q(t_1,t_1)$. A fundamental difference between the two models can be observed at large times. Indeed, for $t_2-t_1 \gg 1$,  $\delta Q(t_1,t_2)$ converges to a $t_1$ independent asymptotic model  $\delta Q_{\infty}<0$. In particular, for  $t_1>t_{st}$ a costant  
heat-flux $\delta Q(t_1,t_2)$ is observed for all values of $t_2$ indicating a stationary non-equilibrium condition. This is confirmed by the parametric plot 
$T R$ vs $C$ (lower panel) indicating that    
curves reach a stationary asymptotic master curve, different from the equilibrium FDT, with $T_{\textrm{eff}}>T$.

%%%%%%%%%%%%%%%%%%%%%%%%%%%%%%%%%%%%%%%%%%%%%%%%%%%%%%%%%%%%%
%%%%%%%%%%%%%%%%%%%%%%%%%%%%%%%%%%%%%%%%%%%%%%%%%%%%%%%%%%%%%

In conclusion, deriving a generalized Harada-Sasa formula~(\ref{J})
from a nonequilibrium FDT we have evidenced its validity for transient
regimes and for systems evolving with jumps involving small entropy changes. 
This justifies the use of the HS relation in the framework of aging systems,
as recently done with experimental results~\cite{GSPC12}.
Moreover, Eq.~(\ref{J}) includes two times rather than one, and can thus be used
to define a generalized heat exchange and an effective temperature in terms of  
quantities that are experimentally accessible. 
It should be interesting to study the existence of similar
relationships in processes with memory~\cite{DN06,MSVW13}, where the
Markov hypothesis is violated, and in the presence of non-linear
contributions in the external field~\cite{LCSZ08}.

\paragraph{Acknowledgments:} We thank C. Maes and A. Puglisi for useful discussions. E.L. acknowledges financial support from MIUR-FIRB RBFR081IUK (2008).
The work of AS is supported by the Granular Chaos
project, funded by the Italian MIUR under the grant number RIBD08Z9JE.

\bibliography{fluct.bib}

\end{document}